\documentstyle[aps,epsfig,multicol]{revtex} 

\def\breakon{\end{multicols}\widetext\vspace{-.2cm}
\noindent\rule{.48\linewidth}{.3mm}\rule{.3mm}{.3cm}\vspace{.0cm}}

\def\breakoff{\vspace{-.2cm}
\noindent
\rule{.52\linewidth}{.0mm}\rule[-.27cm]{.3mm}{.3cm}\rule{.48\linewidth}{.3mm}
\vspace{-.3cm}
\begin{multicols}{2}
\narrowtext}

\begin{document}

\draft

\widetext

\title{Tail States in a Superconductor with Magnetic Impurities}

\author{A Lamacraft and B D Simons} 

\address{Cavendish Laboratory, Madingley Road, Cambridge CB3\ OHE, UK}

\date{\today}

\maketitle 

\begin{abstract}
A field theoretic approach is developed to investigate the profile and 
spectrum of sub-gap states in a superconductor subject to a weak magnetic 
impurity potential. Such states are found to be associated with inhomogeneous 
supersymmetry broken instanton configurations of the action. 
\end{abstract}

%\pacs{PACS numbers: 74.62.Dh, 71.55.-i, 74.40.+k}

%71.55.-i Impurity and defect levels
%73.50.Bk General theory, scattering mechanisms
%74.20.-z Theories and models of superconducting state
%74.40.+k Fluctuations (noise, chaos, non-equilibrium superconductivity,
%localisation, etc.)
%74.62.Dh Effects of crystal defects, doping and substitution

\begin{multicols}{2}

\narrowtext

While spectral properties of an $s$-wave superconductor are largely 
unaffected by weak non-magnetic impurities~\cite{anderson}, the pair-breaking
effect of magnetic impurities leads to the destruction of 
superconductivity. Remarkably, the suppression of the quasi-particle energy 
gap is more rapid than that of the superconducting order parameter, 
admitting the existence of a narrow `gapless' superconducting phase~\cite{ag}. 
According to the conventional (mean-field) theory due to Abrikosov and 
Gor'kov (AG), a gap is maintained up to a critical concentration 
of magnetic impurities. Yet, being unprotected by the Anderson theorem, it is
clear that the gap structure predicted by the mean-field theory is untenable 
and will be destroyed by rare configurations of the random impurity potential. 
Indeed, since the pioneering work of AG, several 
authors~\cite{yl,shiba,rus,makires,bt,bna} 
have explored the nature of the `sub-gap' states. The aim of this work
is to investigate quantitatively the spectrum and profile of 
sub-gap states in weakly disordered superconductors. 

In the earliest works~\cite{yl,shiba,rus}, attention was focussed on the 
the influence of strong impurities. In the unitarity limit, it was shown that
a single magnetic impurity leads to the local suppression of the order
parameter and creates a bound sub-gap quasi-particle state. For a finite 
impurity concentration, these intra-gap states broaden into a band. 
This mechanism contrasts with the AG theory~\cite{ag} for {\em weak} 
magnetic impurities which predicts a gradual suppression of the quasi-particle energy gap. Defining 
$\zeta\equiv 1/\tau_s|\Delta|$, where $|\Delta|$ represents the 
self-consistent order parameter, and $\tau_s$ denotes the Born scattering 
time due to magnetic impurities, one finds $E_{\rm gap}(\tau_s)=|\Delta|(1-
\zeta^{2/3})^{3/2}$, showing an onset of the gapless region when $\zeta=1$
(note $\hbar=1$ throughout). However, even for weak disorder, optimal 
fluctuations of the random potential generate sub-gap states. Extending the 
arguments of Balatsky and Trugman~\cite{bt}, a fluctuation of the random 
potential which leads to an effective scattering rate $1/\tau_s^\prime$ in 
excess of $1/\tau_s$ over a range in excess of the coherence length, 
$\xi=(D/|\Delta|)^{1/2}$, induces quasi-particle states down to energies 
$E_{\rm gap}(\tau_s^\prime)$. These sub-gap states are localised, bound to 
the region or `droplet' where the scattering rate is large, and coupled 
through the proximity effect to the rest of the environment.

The situation bares comparison with band tail states in semi-conductors.
Here rare or optimal configurations of the random impurity potential generate 
bound states, known as Lifshitz tail states~\cite{lifshitz}, which extend
well below the band edge. However, the 
correspondence is, to some extent, superficial: band tail states in 
semi-conductors are typically associated with smoothly varying, nodeless 
wavefunctions. By contrast, the tail states below the superconducting gap 
involve the superposition of states around the Fermi level. As such, one 
expects these states to be rapidly oscillating on the 
scale of the Fermi wavelength, but modulated by an envelope which is localised 
on the scale of the coherence length. This difference is not incidental.
Firstly, unlike the semi-conductor, one expects the spectrum of the 
tails states within the superconducting gap to be `universal', independent of 
the nature of the weak impurity distribution, but dependent only on the 
scattering time $\tau_s$. 
%Secondly, the tail states within each individual droplet will exhibit 
%properties reminiscent of a single {\em isolated} grain or dot coupled 
%through the proximity effect to the rest of the environment. 
Secondly, as we will see, one can not expect a straightforward 
extension of existing theories~\cite{lifshitz,halperin,cardy} of the 
Lifshitz tails to describe the profile of tail states in the 
superconductor.

The aim of this letter is two-fold: firstly we will propose a general field 
theoretic scheme which accommodates the AG theory as a mean-field, and whose
fluctuations determine phase coherence effects in the gapless system. Secondly,
we will relate the appearance of tail states to inhomogeneous supersymmetry 
breaking instanton configurations of the action. 

To keep our discussion simple, we will take the quenched distribution of 
magnetic impurities to be `classical' and non-interacting throughout. For 
practical purposes, this entails the consideration of structures where 
both the Kondo temperature~\cite{makires} and, more significantly, the RKKY 
induced spin glass temperature~\cite{lmk} is smaller than the relevant 
energy scales of the superconductor. The remaining energy scales 
are arranged in the quasi-classical and dirty limits: $\epsilon_F\gg 1/\tau
\gg (|\Delta|,1/\tau_s)$ where $1/\tau$ represents the scattering rate 
associated with non-magnetic impurities. The random system we consider is 
specified by the Gor'kov Hamiltonian 
\begin{equation} \label{eq:hamiltonian}
\hat{H}=\left({\hat{\bf p}^2\over 2m}-\epsilon_F+V({\bf r})\right)
\sigma_3^{\sc ph}+|\Delta|\sigma_2^{\sc ph}+J{\bf S}\cdot{\bf \sigma}^{\sc sp}
\end{equation} 
where Pauli matrices $\{\sigma_i^{\sc ph}\}$ and $\{\sigma_i^{\sc sp}\}$ 
operate on particle/hole and spin indices respectively. Here 
${\bf S}({\bf r})$ and $V({\bf r})$ represent Gaussian $\delta$-correlated 
magnetic and non-magnetic impurity potentials with zero 
mean and variance $\left\langle JS_\alpha({\bf r}) JS_\beta({\bf r}^\prime)
\right\rangle_S=(6\pi\nu\tau_s)^{-1}\delta^d({\bf r}-{\bf r}^\prime)
\delta_{\alpha\beta}$ and $\left\langle V({\bf r}) V({\bf r}^\prime)
\right\rangle_V=(2\pi\nu\tau)^{-1}\delta^d({\bf r}-{\bf r}^\prime)$ 
respectively, and $\nu$ represents the average density of states (DoS).

Phase coherence properties of the Hamiltonian~(\ref{eq:hamiltonian}) rely on 
its invariance under symmetry transformations. The magnetic impurity 
potential breaks both time-reversal and spin rotation symmetry leaving only
charge conjugation symmetry,
\begin{equation} \label{eq:phsym}
\hat{H}=-\sigma_2^{\sc ph}\otimes\sigma_2^{\sc sp}\hat{H}^T
\sigma_2^{\sc sp}\otimes\sigma_2^{\sc ph}.
\end{equation}
When applied to the corresponding Gor'kov Green function, 
$\hat{G}_+(\epsilon)=-\sigma_2^{\sc ph}\otimes\sigma_2^{\sc sp}
\hat{G}_-^T(-\epsilon)\sigma_2^{\sc sp}\otimes\sigma_2^{\sc ph}$, the 
same transformation converts an advanced function into a retarded function 
implying single-particle interference effects as $\epsilon\to 0$~\cite{az}. 
To investigate their influence on spectral (and transport) properties of 
the microscopic Hamiltonian it is convenient to cast the problem in the 
form of a functional field integral. Following a standard 
route~\cite{efetov}, the Gor'kov Green function can obtained from the 
generating function
\begin{eqnarray*}
{\cal Z}[0]=\int D(\bar\Psi,\Psi) \exp\left[i\int d{\bf r}
\bar\Psi(\hat{H}-\epsilon\sigma_3^{\sc cc})\Psi\right],
\end{eqnarray*}
where $\Psi$ represent $16$-component superfields with indices referencing 
the particle/hole ({\sc ph}), spin ({\sc sp}), charge-conjugation ({\sc cc}), 
and boson/fermion ({\sc bf}) spaces, and the Pauli matrices 
$\{\sigma_i^{\sc cc}\}$ act in the {\sc cc} space. The elements of the 
superfields $\Psi$ are not independent and exhibit the symmetry relation 
$\Psi=-\sigma_2^{\sc ph}\otimes\sigma_2^{\sc sp}\gamma \,\bar\Psi^T$ with 
$\gamma={\rm diag}(i\sigma_2^{\sc cc}, -\sigma_1^{\sc cc})_{\sc bf}$. 
%(The consideration of higher point functions demands the further extension 
%of the field space.)

As with normal disordered conductors~\cite{efetov}, when subject to an 
ensemble average over the random impurity distribution, 
the functional integral over the $\Psi$ fields can be traded for
an integral involving matrix fields $Q$ weighted by a non-linear 
$\sigma$-model action. Physically, the fields $Q$, which vary slowly on the 
scale of the mean-free path $\ell$, describe the soft modes of density 
relaxation. When perturbed by the superconducting order parameter, the 
extension of the field integral approach is 
straightforward~\cite{Oppermann,ast}. Taking into account the magnetic 
impurity potential, the local single-particle DoS can 
be expressed as $\langle\nu({\bf r};\epsilon)\rangle_{V,S}=\nu\langle {\rm tr}
(\sigma_3^{\sc ph}\otimes\sigma_3^{\sc cc}\;Q({\bf r}))\rangle_Q/16$, where 
$\langle\cdots\rangle_Q=\int_{Q^2=\openone} DQ\; (\cdots) e^{-S[Q]}$ with
\breakon
\begin{equation} \label{eq:sigmod}
S[Q]=-\frac{\pi\nu}{8}\int d{\bf r}\; {\rm str} \left[D(\partial Q)^2-4
\left(i\epsilon\sigma_3^{\sc cc}\otimes\sigma_3^{\sc ph}-|\Delta|
\sigma_1^{\sc ph}\right)Q-\frac{1}{3\tau_s}(\sigma_3^{\sc ph}\otimes
{\bf \sigma}^{\sc sp}Q)^2\right]\;.
\end{equation}
\breakoff
\noindent
Here $D=v_F^2\tau/d$ represents the classical diffusion constant associated 
with the non-magnetic impurities. The supermatrix fields
are subject to the auxiliary symmetry condition: $Q=\sigma_1^{\sc ph}
\otimes\sigma_2^{\sc sp}\gamma Q^T\gamma^{-1}\sigma_2^{\sc sp}\otimes
\sigma_1^{\sc ph}$. Although the soft mode action is stabilised by the large 
parameter $\epsilon_F\tau\gg 1$, the majority of field fluctuations of the 
action are rendered massive: both the order parameter and the magnetic 
impurity potential lower the symmetry of the low-energy theory. To assimilate 
the effect of these terms, and to establish contact with the AG theory, it 
is necessary to explore the saddle-point equation. 

Varying the action with respect to 
fluctuations of $Q$, subject to the non-linear constraint, one obtains the 
saddle-point or mean-field equation,
\begin{eqnarray*}
&&D \partial\left(Q\partial Q\right) + \left[Q,i\epsilon\sigma_3^{\sc cc}
\otimes\sigma_3^{\sc ph}-|\Delta|\sigma_1^{\sc ph}\right]\nonumber\\
&&\qquad \qquad +\frac{1}{6\tau_s}\left[Q,\sigma_3^{\sc ph}\otimes
{\bf \sigma}^{\sc sp} Q\sigma_3^{\sc ph}\otimes
{\bf \sigma}^{\sc sp}\right] = 0\;.
\end{eqnarray*}
With the ansatz: $Q_{\sc mf}=\sigma_3^{\sc cc}\otimes\sigma_3^{\sc ph}
\cosh\hat\theta+i\sigma_1^{\sc ph}\sinh\hat\theta$, where $\hat\theta=
{\rm diag}(\theta_1,i\theta)_{\sc bf}$, the saddle-point equation decouples
into boson and fermion sectors and takes the form
\begin{eqnarray}
\partial_{{\bf r}/\xi}^2\hat\theta+2i\left(\cosh\hat\theta-\frac{\epsilon}
{|\Delta|}\sinh\hat\theta\right)-\zeta\sinh(2\hat\theta)=0 \;,
\label{eq:usadel}
\end{eqnarray}
a result reminiscent of the Usadel equation of quasi-classical 
superconductivity~\cite{usadel}. This is no coincidence: when 
subject to an inhomogeneous order parameter, the same effective 
action~(\ref{eq:sigmod}) describes the proximity effect in a hybrid 
normal/superconducting compound~\cite{ast}. In the present context, when 
combined with the
self-consistent equation for the order parameter~\cite{footnote}, the 
homogeneous form of Eq.~(\ref{eq:usadel}) coincides with the mean-field 
equations obtained by AG~\cite{ag,footnote_bna}. 

The AG solution is not unique: for $\epsilon\to 0$, 
the saddle-point equation admits an entire manifold of homogeneous solutions 
parameterised by the transformations $Q=TQ_{\sc mf}T^{-1}$ where 
$T=\openone_{\sc ph}\otimes\openone_{\sc sp}\otimes t$ and $t=\gamma
(t^{-1})^T\gamma^{-1}$: soft fluctuations of the fields, which are controlled 
by a non-linear $\sigma$-model defined on the manifold $T\in {\rm OSp}(2|2)/
{\rm GL}(1|1)$ (symmetry class D in the classification of 
Ref.~\cite{zirn_class}), control the low-energy, long-range 
properties of the gapless system giving rise to unusual localisation and 
spectral properties. (For a comprehensive discussion of 
the physics of the gapless phase, we refer to 
Refs.~\cite{az,bcsz2,sf,rg,bsz,ls}.)

This completes the formal description of the bulk superconducting phase. The
mean-field solution of the AG equation defines the 
global phase structure of the bulk states. Soft fluctuations around the 
AG mean-field describe phase coherence effects due to quantum interference. 
However, within the present scheme it is not yet clear 
how to accommodate sub-gap states in the gapped phase of the 
AG theory. To identify such states, it is necessary to return to the 
saddle-point equation~(\ref{eq:usadel}) and seek spatially {\em inhomogeneous} 
solutions. We will see that such configurations necessarily break 
supersymmetry.

To keep our discussion simple, let us focus on a quasi one-dimensional 
geometry. The generalisation to higher dimensions follows straightforwardly.
To stay firmly within the diffusive regime, we require the system size $L$ to 
be much smaller than the localisation length of the normal system 
$\xi_{\rm loc.}\simeq \nu L_{\rm w} D$, where $L_{\rm w}$ denotes the 
cross-section. Furthermore, we focus on the interval near to the gapless 
region (i.e. $\zeta\lesssim 1$), where self-consistency of the order 
parameter can be safely neglected. 
To define the inhomogeneous field configurations it is convenient to 
deal not with the saddle-point equation~(\ref{eq:usadel}) itself, but its
first integral, $(\partial_{x/\xi}\hat\theta)^2+V(\hat{\theta})={\rm const}.$, 
where 
\begin{eqnarray*}
V(\hat\theta)\equiv 4i\left(\sinh\hat\theta-\frac{\epsilon}{|\Delta|}\cosh
\hat\theta\right)-\zeta\cosh (2\hat\theta) \;,
\end{eqnarray*}
represents the effective complex `potential'. Let us denote by 
$\theta_{\sc ag}$ the 
values of $\theta_1$ and $i\theta$ at the conventional AG saddle-point. From 
the mean-field DoS, $\nu(\epsilon)=\nu\;{\rm Re}\cosh\theta_{\sc ag}$, it is 
evident that for $\epsilon<E_{\rm gap}$, ${\rm Im}\;\theta_{\sc ag}=\pi/2$. 
The corresponding value of ${\rm Re}\;\theta_{\sc ag}$ depends sensitively 
on the energy, with ${\rm Re}\;\theta_{\sc ag}=0$ for $\epsilon=0$. Taking 
into account the condition that the solution should coincide with 
$\theta_{\sc ag}$ at infinity, one can identify a ``bounce'' 
solution parameterised by $\theta_1=i\pi/2+\phi$, with $\phi$ real, 
involving the \emph{real} potential $V_{\sc r}(\phi)\equiv V(i\pi/2+\phi)$ 
with endpoint $\phi'$ such that $V_{\sc r}(\phi^\prime)=V_{\sc r}
(\phi_{\sc ag})$.

\begin{figure}[hbtp]
\begin{center}
\includegraphics[width=6cm]{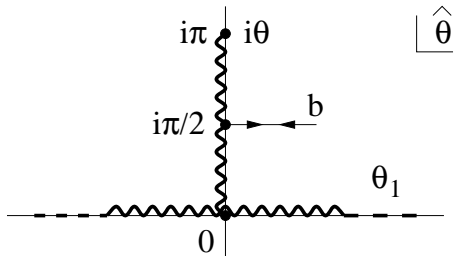}
\caption{Integration contours for boson-boson and fermion-fermion fields in 
the complex $\hat\theta$ plane. The bounce solution for $\epsilon=0$ 
(labelled as `b') is shown schematically.}
\label{fig:bounce_contour}
\end{center}
\end{figure}

Now integration over the angles $\hat{\theta}$ 
is constrained to certain contours~\cite{efetov}. Is the bounce solution 
accessible to both? As usual, the contour of integration over the boson-boson 
field $\theta_1$ includes the entire real axis, while for the fermion-fermion 
field $i\theta$ runs along the imaginary axis from $0$ to $i\pi$. With a 
smooth deformation of the integration contours, the AG saddle-point is 
accessible to both the angles $\hat{\theta}$~\cite{ast}. By contrast, the 
bounce solution {\em and} the AG solution can be reached simultaneously by a 
smooth deformation of the integration contour {\em only} for the boson-boson 
field $\theta_1$ (see Fig.~\ref{fig:bounce_contour}). The bounce is associated 
with a {\em spontaneous breaking of supersymmetry} at the level of the saddle 
point. 

Having identified the saddle-point field configuration, we now turn to the
role of fluctuations. Here we sketch the important 
aspects of the expansion, referring to Ref.~\cite{ls} for a detailed analysis.
Field fluctuations can be separated into ``radial'' and ``angular'' 
contributions. The former involve fluctuations of the diagonal elements 
$\hat{\theta}$, while the latter describe rotations including those Grassmann 
transformations which mix the {\sc bf} sector. Both classes of fluctuations 
play a crucial role.

As usual, associated with radial fluctuations around the bounce, there 
exists a zero mode and a negative energy mode due to translational invariance 
of the solution. The latter, which necessitates a $\pi/2$ rotation of the 
corresponding integration contour to follow the line of steepest descent (c.f. 
Ref.~\cite{coleman}), has two effects: firstly it ensures that 
the non-perturbative contributions to the local DoS are non-vanishing, and 
secondly, that they are positive. Turning to the angular fluctuations, the 
spontaneous breaking of supersymmetry is accompanied by the appearance of a 
Grassmann zero mode 
%in the {\sc bf} sector 
separated by a gap from higher 
excitations. The zero mode 
ensures that the supersymmetry breaking saddle-point respects 
the normalisation condition $\langle{\cal Z}[0]\rangle_{V,S}=1$, and that 
the local DoS is non-vanishing only in the vicinity of the bounce.

\begin{figure}[hbtp]
\begin{center}
\includegraphics[width=7.5cm]{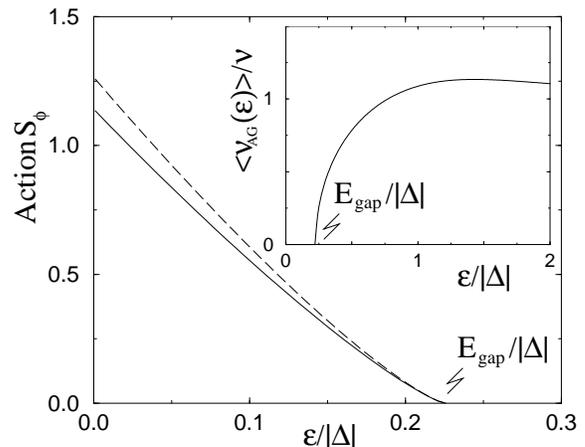}
\caption{Action $S_\phi$ for $\zeta=0.5$ obtained numerically (solid curve) 
together with the expansion in $E_{\rm gap}-\epsilon$ (dotted curve) as 
determined by Eq.~(\ref{action_expansion}). Note that the action vanishes as 
$\epsilon\to E_{\rm gap}$. The AG solution for the DoS is shown inset.}
\label{fig:dos}
\end{center}
\end{figure}

Taking into account Gaussian fluctuations and zero modes, one obtains the 
non-perturbative, one instanton contribution to the sub-gap DoS:
\begin{eqnarray} \label{eq:dosresult}
&&{\langle\nu(\epsilon)\rangle_{V,S}\over\nu}\sim\; (-i|K|)\int dx\;
i(\sinh\phi(x)-\sinh\phi_{\sc ag})|\varphi_0(x)|^2\nonumber \\
&&\qquad\qquad\times \sqrt{LS_\phi\over\xi}\; \exp\left[-4\pi\nu L_{\rm w}
\sqrt{D|\Delta|} S_\phi\right]\;,
\end{eqnarray}
where $S_\phi\equiv \int_{\phi_{\sc ag}}^{\phi'} d\phi 
\sqrt{V_{\sc r}(\phi_{\sc ag})-V_{\sc r}(\phi)}$.
%
%\begin{eqnarray*} 
%S_\phi\equiv \int_{\phi_{\sc ag}}^{\phi'} d\phi \sqrt{
%V_{\sc r}(\phi_{\sc ag})-V_{\sc r}(\phi)}\;.
%\end{eqnarray*}
%
Here, the factor $\sqrt{L S_\phi/\xi}$ represents the Jacobian 
associated with the introduction of the collective coordinate~\cite{coleman}, 
$-i|K|$ is the overall factor from the non-zero modes, and $\varphi_0(x)$
represents the normalised eigenfunction associated with the Grassmann zero 
mode~\cite{ls}. Fig.~\ref{fig:dos} shows the action $S_\phi$ for a typical
value of $\zeta$. These results show that, for arbitrarily small but finite 
impurity concentrations, the DoS remains {\em finite} even at $\epsilon=0$.
For energies $\epsilon$ just below $E_{\rm gap}$, an analytical solution can
be obtained for the bounce in arbitrary dimension. Generally, one finds 
the exponent $4\pi g(\xi/L)^{d-2} S_\phi$ where
\begin{eqnarray}
S_\phi=a_d\ \zeta^{-2/3} (1-\zeta^{2/3})^{-(2+d)/8}\left({E_{\rm gap}-
\epsilon\over |\Delta|}\right)^{(6-d)/4}.
\label{action_expansion}
\end{eqnarray}
Here $g=\nu D L^{d-2}$ denotes the bare conductance, and $a_d$ 
is a numerical constant ($a_1=8\ {}^4\sqrt{24}/5$). Furthermore, the optimal 
solution extends over a length scale $\xi ((E_{\rm gap}-\epsilon)/
|\Delta|)^{-1/4}$, set by the coherence length (as expected), and diverges
as $\epsilon\to E_{\rm gap}$ (c.f. Ref.~\cite{lo}).

To conclude, let us make some remarks. Firstly, the procedure outlined above 
has a number of close relatives in the literature. As well as the 
investigation of tail states in semi-conductors~\cite{cardy}, a
supersymmetric field theory was developed by Affleck~\cite{affleck} (see 
also Refs.~\cite{em}) to investigate tail states in the lowest Landau level. 
There it was shown that tail states correspond to supersymmetry broken  
configurations of the {\em $\Psi$-field action} (c.f. Ref.~\cite{cardy}). It 
is also interesting to compare the present scheme with the study of 
`anomalously localised states'~\cite{km} (see also, Ref.~\cite{fe}). There 
one finds that long-time current relaxation in a disordered wire is also 
associated with spontaneous breaking of supersymmetry. Finally, we note 
that the Lifshitz argument has been applied on the level of the Usadel 
equation in the study of gap fluctuations due to inhomogeneities of the
BCS interaction~\cite{lo}.

Although we have focussed on the question of tail states in the 
superconducting gap, the general scheme is more widely 
applicable. For example, in the present system, the transition to bulk 
superconductivity will be preempted by the nucleation of superconducting 
islands or droplets within the metallic/insulating phase (c.f. 
Ref.~\cite{ioffe}). Similarly, the Stoner transition to a bulk 
itinerant ferromagnet in a disordered system will be mediated by the 
formation of a droplet phase in which islands become 
ferromagnetic~\cite{aleiner}. In both cases, we expect these `droplet 
phases' to be associated with inhomogeneous (replica) symmetry broken 
saddle-point field configurations of the corresponding low-energy action. 

{\sc Acknowledgements}: We are grateful to Alexander Altland, Alexander 
Balatsky, Alex Kamenev, and Mike Stone for valuable discussions. We are also 
deeply indebted to Dima Khmel'nitskii for bringing this general subject to our 
attention, and to John Chalker for crucial discussions, particularly those 
made at an early stage of this work. One of us (AL) would like to acknowledge 
the financial support of Trinity College.

\end{multicols}


\begin{references}
\bibitem{footnote_bna} For a random matrix Hamiltonian of rank $N$, 
corrections to the homogeneous AG equation from massless fluctuations enter 
at $O(N^{-1})$ implying that the density of tail states vanishes in the 
limit $N\to\infty$. We note that this conclusion contradicts the findings of 
Ref.~\cite{bna}.
\end{references}
\end{document}